# *Glassy behavior of interface states in Al-AlOx-Al tunnel junctions*

Jeremy R. Nesbitt and Arthur F. Hebard

*Department of Physics, University of Florida, Gainesville, FL 32611-8440*

Abstract:

We present results of a study of tunnel junction aging in which the early time dynamics are captured by *in situ* monitoring of electrical properties of Al-AlOx-Al planar tunnel junctions beginning when the deposition of the counterelectrode is complete. The observed stretched exponential dependences of the conductance and the capacitance manifest hierarchically constrained dynamics imposed by correlated relaxations of interface traps. Bias voltage is used as a control parameter to create bias-dependent aging trajectories that exhibit memory and age-dependent relaxations. Simple tunnel barrier and equivalent circuit modeling provide a comprehensive understanding of this novel and unexpected glassy behavior that appears to be unique to tunnel junctions and has important implications for technical applications.

The practice of reliably fabricating high quality tunnel-junction barriers for electronic applications depends on the choice of barrier material, the compatibility of the electrodes sandwiching the barrier and the properties of the interfaces. Emerging technologies relating to magnetoelectronics[1] and cryogenic sensors on a chip[2] together with developing opportunities in fundamental nanoscience research[3] require high quality metal oxide barriers. As an example, aluminum is known to completely wet transition metal surfaces and, when oxidized to completion, forms a tunnel barrier suitable for magnetic[4,5] or Josephson [6] tunnel junctions. Aluminum oxide (AlOx) barriers are well studied and there is recognition that the deleterious presence of a broad distribution of electronic states both in the oxide and at the metal/metal oxide interfaces can provide additional conducting paths[7,8], be the source of noise[9] and compromise breakdown strength[10,11]. Accordingly, a common goal of many researchers is to develop a comprehensive understanding of the generation, interaction and neutralization of interface traps[4,8,11,12]. Our perspective on this problem has been influenced by *in situ* observations of the time evolution of the electrical properties of Al-AlOx-Al tunnel junctions beginning at the time the deposition of the counterelectrode is completed. We have found that the time evolution of concurrent changes in the conductance and the capacitance (i.e., aging) is most pronounced at early times. During aging, we have also studied the driven off-equilibrium dynamics by perturbing the system with dc bias voltages. Using these techniques, the presence of bias-dependent aging trajectories (lifelines), a decreasing responsiveness (reaction time) with increasing age, and memory effects in both the conductance and the capacitance lifelines become apparent.

The microscopic mechanisms underlying this glassy behavior are related to the generation, correlated interaction, and annealing of interface states or traps. When an electron tunnels or is thermally activated into or out of an interface state, a simultaneous change in the coordination of atoms occurs in the vicinity of the trap. These changes in charge distribution, which could also be caused by mobile oxygen vacancies[13], give rise to changes in the tunnel barrier height[14], which in turn affect neighboring traps. For thin tunnel junction barriers the conductance and capacitance are respectively dominated by exponential sensitivity to tunnel barrier parameters[15] and a *thickness-independent* interface capacitance[16, 17]. Accordingly, the strongly correlated behavior of electronic motion and relaxation of electric dipoles associated with trapped charge give rise to the observed temporal changes and directly derive from the tunnel coupling of the two metal/metal-oxide interfaces defining the tunnel junction. In this context, it is appropriate to refer to this glassy system as a tunnel-junction glass (TJG). The TJG has much in common with other glassy systems that manifest their properties in phenomenology such as spin relaxation[18], structural relaxation in polymers[19], electric-field-gated resistance[20], gate controlled charge injection[21], photo-stimulated resistance[22] and dielectric response[23]. Our semi-quantitative analysis, based on simple tunneling concepts[7, 15, 24], provides a comprehensive understanding of the glassy behavior that is observed and at the same time accentuates the critical role of the microscopic relaxation processes associated with the interfaces.

The *in situ* measurements were performed in a vacuum deposition chamber with a base pressure below $5\times10^{-7}$ Torr. The Al base electrodes were deposited through shadow masks onto precleaned glass substrates. The AlOx barrier was then formed in a dc glow discharge (100 mTorr oxygen) for 1-10 minutes without breaking vacuum. Finally, the Al counterelectrode was deposited as a cross stripe through an electrically isolated mask. Junction areas are $8.9 \times 10^{-3}$ cm$^2$. Electrical measurements commenced when the counterelectrode made contact with the predeposited contact pads and a tunneling current was detected. This usually occurred when the counterelectrode reached a thickness of ~200 Å. The 4-terminal complex admittance was measured using phase sensitive detection at frequencies 1 Hz – 10 kHz with typical ac rms excitations of 50 mV. This excitation level was confirmed to be at a low enough level to assure linearity. All measurements were made *in situ* at room temperature and at system base pressure. Lower temperatures led to a considerable slowing down of the relaxations thus highlighting the relevance of thermally activated processes.

The simultaneous measurement of the time-dependent conductance and capacitance shown in Fig. 1 is motivated by the recognition that the admittance rather than impedance is more suited to obtaining a physically meaningful model of tunnel junction[7] and thin-film oxide behavior[12]. Accordingly, we model our tunnel junction response (see Fig. 1 inset) as a parallel combination of a dc tunnel resistance $R_0$ and a complex frequency-dependent capacitance $C^*(t,\omega) = C_1(t,\omega) - iC_2(t,\omega)$. Both $R_0$ and $C^*(t,\omega)$ in this parallel $R_0C^*$ model are influenced by interface processes: $R_0$ because of its exponential sensitivity to $d\phi^{1/2}$, where $\phi$ is the barrier height and $d$ the electrode spacing, and $C^*(t,\omega)$ because of the dominance of interface capacitance over the series-connected geometrical capacitance[7, 17].

We emphasize the distinction between the short time scales, $\omega^{-1}$, associated with each measurement and the much longer times, $t$, associated with aging of the junction.

With this partitioning of time dependences, we implicitly assume that any given measurement is made on a fast enough time scale to assure that changes associated with aging can be safely ignored. Although this parallel $R_0C^*$ model is overly simplified, it does capture the behavior of the aging trends shown in Fig. 1. Both the conductance, $G(t,\omega) = 1/R_0 + \omega C_2(t,\omega)$, which is measured at 1 Hz so that it approximates a dc measurement, and the capacitance, $C_1(t,\omega)$, which is measured at higher frequencies (13-1389 Hz), are found to have a stretched exponential dependence. Aging primarily affects the real and imaginary parts of $C^*(t,\omega)$ through the annealing of traps. As the junction ages and the traps are annealed or effectively neutralized, there is a simultaneous reduction of trap assisted tunneling processes and polarization at trap sites. Hence the reduction with aging of both $G(t,\omega)$ and $C_1(t,\omega)$ is observed. All of the time dependence is incorporated into the complex capacitance $C^*(t,\omega)$, and $R_0$ represents the dc resistance of a trap-free fully aged ideal tunnel junction. In the fully aged asymptotic state when $t \to \infty$ there is no dispersion associated with the interface, i.e., $C_2(t,\omega) = 0$, and $C_1(t,\omega) = \text{Re}\{C^*(t,\omega)\}$ becomes a frequency-independent constant as required by Kramers-Kronig constraints between the real and imaginary parts of $C^*(t,\omega)$.

The measurement time $t$ is initialized to zero when electrical measurements begin during the counterelectrode deposition. The 'dc' conductance $G_{dc} = G(t)$ measured at 1 Hz is found to obey a stretched exponential dependence of the form $G_{dc}(t) = G_0 \exp[-((t+t_0)/\tau)^\beta] + G_\infty$ where $G_0$ and $G_\infty$ are the initial ($t = 0$) and final state ($t \to \infty$) conductivities, $\tau$ is a characteristic relaxation time and $\beta$ is an exponent that is constrained to lie between zero and one. Similar expressions describe $C_1(t,\omega)$. The junction age $t_{age} = t + t_0$ is defined in terms of an offset time $t_0$ which we find from the fitting procedure to always have a value less than the time interval from the beginning of the counterelectrode deposition to the start of the measurement clock. In other words, the results of the fitting procedure used to determine the five parameters which best describe the stretched exponential dependences give an estimate of the time when the junction is 'born' during the deposition of the counterelectrode. Accordingly, the birthing time is the earliest time marking the onset of meaningful electrical measurements. Independent fits of $G_{dc}$ and $C_1(t,\omega)$ shown in Fig. 1 (four curves) give $t_0 = 56(6)$ sec, $C_\infty = 14.1(1)$ nF and $\beta = 0.12(3)$. The similar values for $t_0$ imply the same birthing time for all components of the admittance, and the less than 1% variation in $C_\infty$ (3 curves) justifies the use of the parallel $R_0C^*$ model with its implied convergence at all frequencies to a *single* $t \to \infty$ steady state value. The relatively low values found for the parameter $\beta$ reflect a broad distribution of relaxation times that give rise to pronounced changes at early time and more gradual changes on extended time scales. We have investigated $G(t)$ in more than thirty junctions having initial resistances ranging from 50 to 200,000 Ω and found that almost half of them adhere well to the stretched exponential form with $\beta = 0.23(8)$ for 14 junctions. Adequate fits could not be obtained if there are insufficient early-time data or if the junctions are of poor quality with presumed resistive shorts. Stretched exponential behavior is generally associated with glassy systems having serial relaxations governed by a hierarchy of constraints[25] and thus seems to be an appropriate description of the correlated relaxations of the traps of the TJG.

Generically, glasses are systems that, when driven out of equilibrium with the application of an external control parameter, evolve slowly toward an equilibrium state that depends on the same control parameter. Shown in Fig. 2 is the effect on resistance $R = 1/G$ when a dc voltage bias $V_b = 0.5$ V is applied. These results can be qualitatively understood by invoking the Simmons model[15] in which electron transport across a barrier of effective height $\phi$ and thickness $d$ gives a conductance $G \propto \exp[-\gamma d \phi^{1/2}]$ where $\gamma = 2\sqrt{2m^*}/\hbar$ and $m^*$ is the effective mass of an electron in the barrier. Voltage dependent terms in $\phi$ lead to an increase in $G$ (decrease in $R$) when $V_b$ of either sign is applied. This dependence explains the switching from a high resistance trajectory at $V_b = 0$ to a lower resistance trajectory (dashed line) at $V_b = 0.5$ V. Resistance $R$ rather than $G$ is plotted because the bias-dependent aging trajectories are more linear, thus clarifying the notion with a guide to the eye that there is a multiplicity of $V_b$-dependent 'lifelines'.

A closer look at the trajectories in Fig. 2 reveals that switching $V_b$ at the times indicated by the vertical dashed lines results in a fast (electronic) change in $R$ followed by a slow (ionic) relaxation to the new lifeline. Analysis of these $V_b$-induced relaxations is shown in Fig. 3 where portions of the $V_b = 0$ lifeline data of Fig. 2 are plotted on double logarithmic axes. The 0.5 V bias is held constant for the times indicated in the legend and the change in resistance distinguishes itself by obeying the relation, $\delta R(t_r) = R_0 t_r^\alpha$, where $t_r$ is the time elapsed from the removal of $V_b$, and $\alpha$ is a power law exponent determined by the slopes shown in the $\delta R$ vs $t_r$ plots of Fig. 3. A similar analysis applied to the relaxations of the lower $V_b = 0.5$ V lifeline data reveals a smaller set of exponents $\alpha'$. The exponents $\alpha$ and $\alpha'$ both decrease with increasing age with a roughly exponential dependence as shown by the solid lines in the inset of Fig. 3.

An insightful qualitative understanding of these trends can be gained by replacing $\phi$ by $\phi + \delta\phi$ in the Simmons expression where $\delta\phi = \phi_b \ln(t_r/\tau_b)$ is a small parameter ($\delta\phi \ll \phi$) defining the time dependence of the $V_b$-induced relaxations in terms of the constants $\phi_b$ and $\tau_b$. Expanding with respect to the ratio $\delta\phi/\phi$ provides us with the Simmons model approximation $G \sim \exp(-\gamma d \phi^{1/2})(t_r/\tau_b)^\alpha$ where $\alpha = \gamma d \phi_b / 2\phi^{1/2}$. In agreement with experiment, both $\alpha$ and $\alpha'$, which are proportional to $\phi^{1/2}$, decrease with increasing age (see Fig. 3 inset), since the increasing tunnel junction resistance $R$ (decreasing $G$) implies a concomitantly increasing $\phi$. We also note that, since $\alpha \propto \phi^{1/2}$, the larger age-dependent $\alpha$ on the high resistance $V_b = 0$ upper lifeline of Fig. 2 has memory of the smaller $\phi$ of the low-resistance $V_b = 0.5$ V lifeline. In like manner, the smaller $\alpha'$ of the low-resistance lifeline has memory of the time spent prior to the $V_b$-induced transition on the high resistance lifeline with its higher value of $\phi$.

The effect of $V_b$ on $\phi$ has also been determined by using 'witness' bias sweeps of the conductance $G(V)$ at different $V_b$-dependent ages to extract barrier parameters. A witness bias sweep requires a brief interruption (~2 min) of the constant $V = V_b$ condition so that a snapshot of $G(V)$ over the full range of voltage (-0.5V $< V <$ 0.5V) can be obtained. Our analysis relies on a theory[24] for asymmetrical tunnel barriers depicted schematically in the inset of Fig. 4. A newly formed junction was allowed to age for 3h at $V_b = 0$, at which time a short witness sweep was made. A bias voltage of 0.5 V was then applied and the junction allowed to age for another 3h before a second witness sweep was applied, etc. In this manner the four rows of Table I were obtained. Two important trends are apparent. Firstly, the parameters $\phi_1$, $\phi_2$ and $d$ all increase steadily with age thus affirming the applicability of the Simmons model and the earlier assumption that an increase of $d\phi^{1/2}$ is connected to a decrease of $G$ with age. The corresponding increase in $d$ should not be taken at face value because there is also a concomitant decrease in shunting paths as the quality of the barrier improves with age. The second important trend to note from Table I is the noticeably larger increase of $\phi_2$ relative to $\phi_1$ for positive $V_b$; whereas the opposite trend occurs for negative $V_b$. This result can be understood by referring again to the inset of Fig. 4, which depicts positive bias and illustrates how the ionic charge in the barrier redistributes on long time scales in response to the applied electric field $E$ to preferentially increase $\phi_2$ relative to $\phi_1$.

These same electronic and ionic motions are also responsible for the $V_b$-induced capacitance changes shown in the main panel of Fig. 4. The three curves represent capacitance witness sweeps after waiting 740 minutes with $V_b = 0.4$ V applied (curve 1), thereafter immediately changing to $V_b = -0.4$ V and waiting an additional 40 minutes (curve 2) and finally another 30 minutes (curve 3) for a total of 70 minutes at $V_b = -0.4$ V. The evolution of the capacitance from the minimum at $V_b = 0.4$ V to a new minimum at $V_b = -0.4$ V while still displaying a memory of its old minimum is reminiscent of 'two-dip' minima seen in low temperature field-gated conductance sweeps of amorphous indium oxide films[26] and the dielectric response at $T = 140$mK of polymer films exposed to fields $E \sim 0.1$MV/cm[23]. A theoretical interpretation of the low-temperature polymer data posits glassy behavior arising from an interacting broad distribution of tunneling two level systems responding to electric field induced strains[27]. The picture is only valid if $E$ due to the application of $V_b$ is large enough to assure $pE > k_BT$ where $p$ is the electric dipole moment associated with a typical trap. Using the 0.8V separation between the capacitance minima in Fig. 4 and assuming $d = 10$ Å ($E \sim 2.5$ MV/cm) we calculate $p = k_B T/E = 1.6$ D, a value which is close to typical dipole moments in glasses[23]. The minimum in the capacitance corresponds to a 'hole' in the distribution of local fields and is only stable (equilibrium) when the number of dipoles experiencing zero local electric field is exactly zero[27]. Otherwise, a dipole could flip without costing energy but surely disturbing the equilibrium. The application of a different bias voltage (Fig. 4) creates additional polarization (capacitance increase), a new hole develops and memory of the former hole lingers.

The microscopic picture that emerges as a basis for understanding the tunnel-junction glass (TJG) involves a complex interplay of fast electronic and slow ionic correlated interactions. With respect to applications, it is reasonable to expect that all metal/oxide interfaces have a finite density of interface states that can become problematic when current leakage, voltage breakdown, time stability (noise), carrier mobility and/or spin-flip scattering is a concern. We believe that the TJG behavior discussed here is a general phenomena, applicable to a wide variety of materials, that manifests itself during the early stages of interface formation but then has sharply attenuated consequences if the junction has aged sufficiently by the time it is taken out of the vacuum system, or if it is immediately cooled to low temperatures. For applications, our results suggest a strategy of using *in situ* monitoring of processing steps to accelerate the aging process and assure a minimum of interface traps. Such strategies might include annealing or flood-gun electron bombardment, as done for high quality AlOx layers[13], or hydrogen exposure to passivate dangling bonds, as done for the Si/SiO$_2$ interface[29]. Despite such palliatives, interface traps are ubiquitous and if created in sufficient numbers can, as we have shown here, give rise to glassy behavior that has fascinating properties in common with other glassy systems, namely: birth, memory, relaxation on extended time scales, externally-influenced lifelines, response times that decrease with increasing age, and an ultimate cessation of all activity as $t \to \infty$.

We are indebted to Kevin T. McCarthy who performed the preliminary experiments on *in situ* tunnel junction aging. We are also grateful to Stephen Arnason, Xu Du, Selman Hershfield, Clare Yu, Michael Weissman and Neil Zimmerman for stimulating discussions and for posing challenging questions. This work is supported by the NSF(DMR) under Grant No. 0404962

.

**Table 1:** Barrier height parameters calculated at different ages and bias voltages from witness bias sweeps of the conductance as described in the text.

| Age (h) | Vb (V) | $\phi_1$ (V) | $\phi_2$ (V) | d (Å) |
|---|---|---|---|---|
| 3 | 0 | 1.06 | 3.55 | 18.8 |
| 6 | 0.5 | 0.995 | 4.92 | 21.1 |
| 9 | 0 | 1.21 | 5.36 | 23.4 |
| 12 | -0.5 | 1.48 | 4.98 | 24.3 |

# Figure Captions

**Figure 1:** Data (symbols) showing the time dependence (age) of the conductance (upper panel) and the capacitance (lower panel) taken at the three frequencies indicated in the inset. The solid lines are stretched exponential fits to the respective curves with fitting parameters given in the text. The tunnel junction admittance is modeled as a dc resistance $R_0$ in parallel with a time-dependent complex capacitance (inset) as discussed in the text. At infinite time all of the dc current flows through $R_0$, and there is no dispersion in the capacitance (i.e., $C_2(\omega) = 0$ and C1 is a constant).

**Figure 2:** Plot of the time dependence of resistance aging trajectories for zero applied dc bias voltage $V_b = 0$ (upper solid curve) and $V_b = 0.5$V (lower dashed curve). Upon switching $V_b$, the transitions between the two lifelines occur on a fast electronic scale (vertical dashed arrows) followed by slow electronic (involving ions) relaxations, which are more pronounced on the upper lifeline.

**Figure 3:** Plots on logarithmic axes of the relative change in resistance $\delta R/R$ as a function of time $t_r$ measured after $V_b$ is held at 0.5V for the times $t_{on}$ indicated in the legend and then switched at $t_r = 0$ to $V_b = 0$. The power law exponents α are equal to the slopes of the solid line fits in the main panel and plotted as solid circles against the sample age in the inset. A similar set of exponents α′, plotted as solid squares using the right hand axis of the inset, are determined from the power-law dependences (not shown)

of the relaxations to the lower lifeline of Fig. 2. Both α and α′ decrease with a roughly exponential dependence on sample age as shown by the solid line fits to the data.

**Figure 4:** Witness voltage sweeps of the capacitance after waiting 740 minutes with $V_b = 0.4V$ applied (curve 1), thereafter immediately changing to $V_b = -0.4V$ and waiting an additional 40 minutes (curve 2) and finally another 30 minutes (curve 3) for a total of 70 minutes at $V_b = -0.4V$. The evolution of the minimum at 0.4V to a new minimum at -0.4V while still retaining a memory of the old minimum manifests memory in the imaginary part of the admittance. The fixed point crossing near 0.1V is consistent with the decomposition of each curve into two independent components. A schematic of the asymmetric barrier with barrier parameters $\phi_1$, $\phi_2$, and $d$ (see Table I) subject to the bias $V_b$ is shown in the inset.

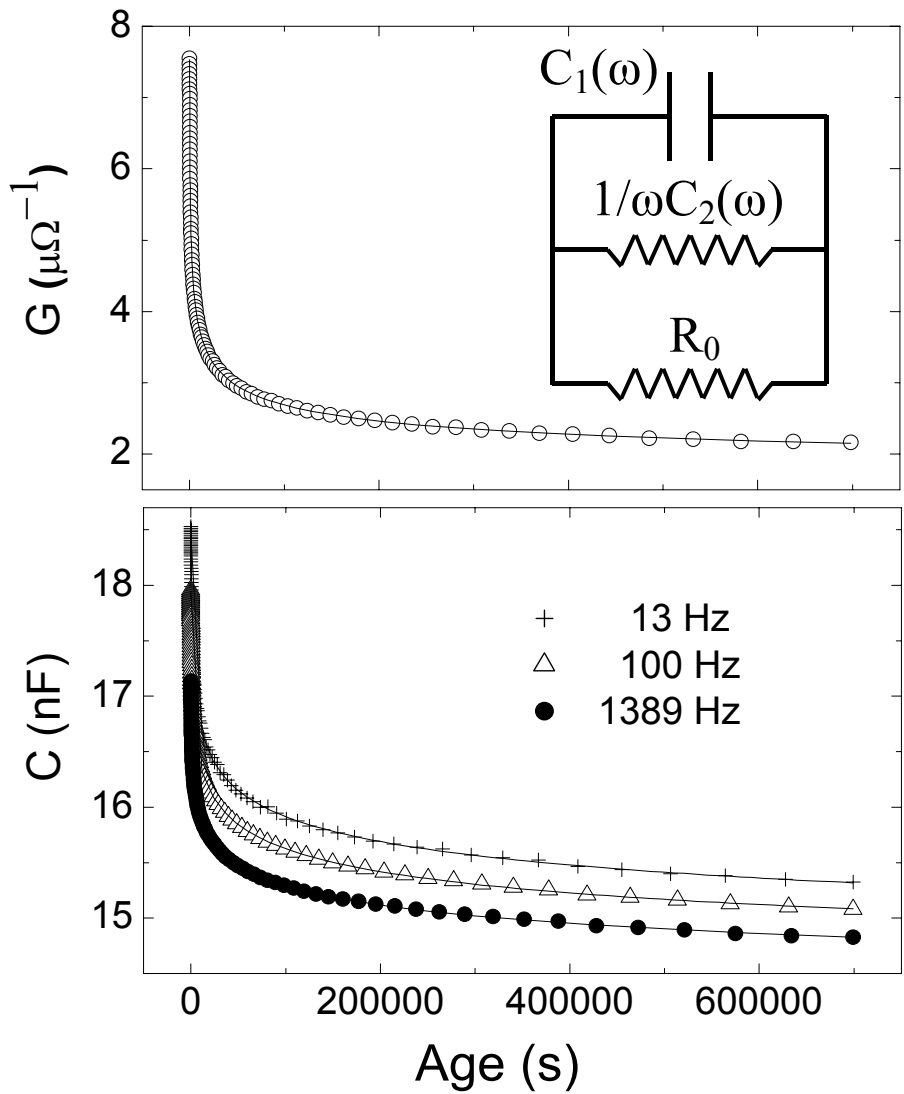

Nesbitt/Figure 1

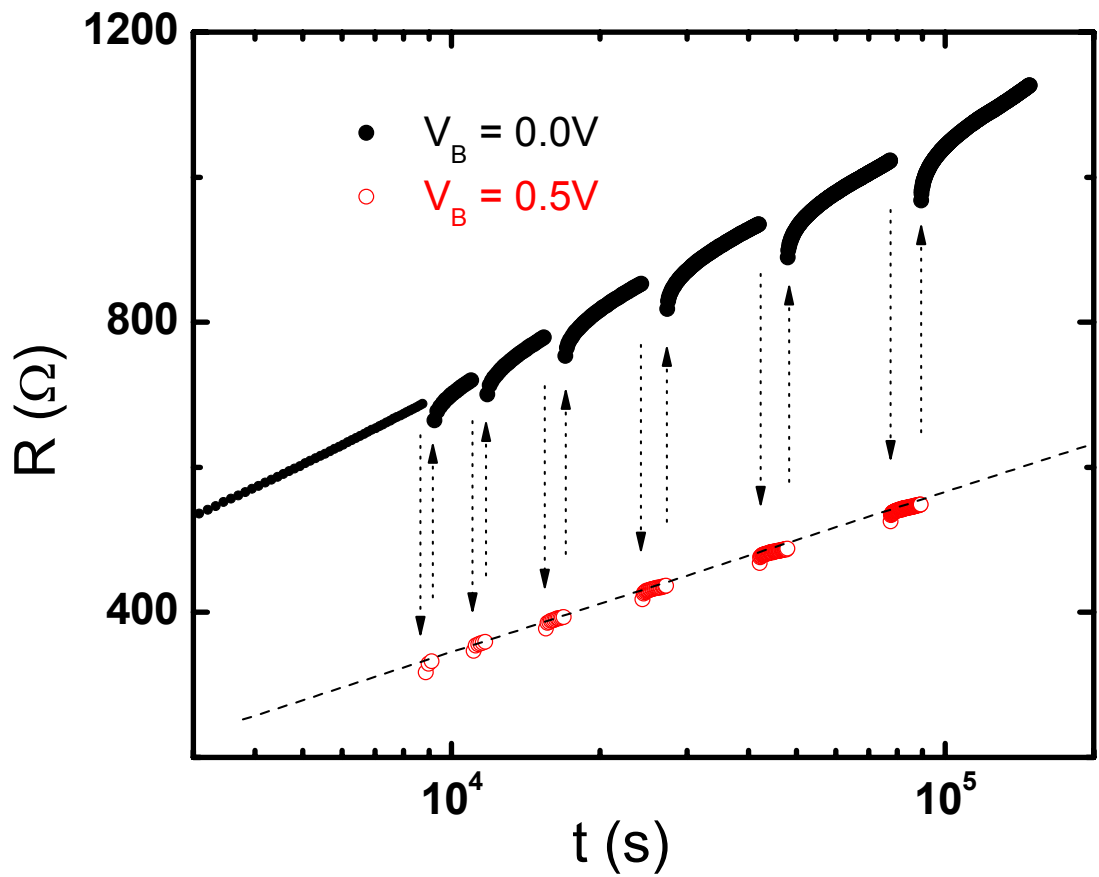

Nesbitt/Figure 2

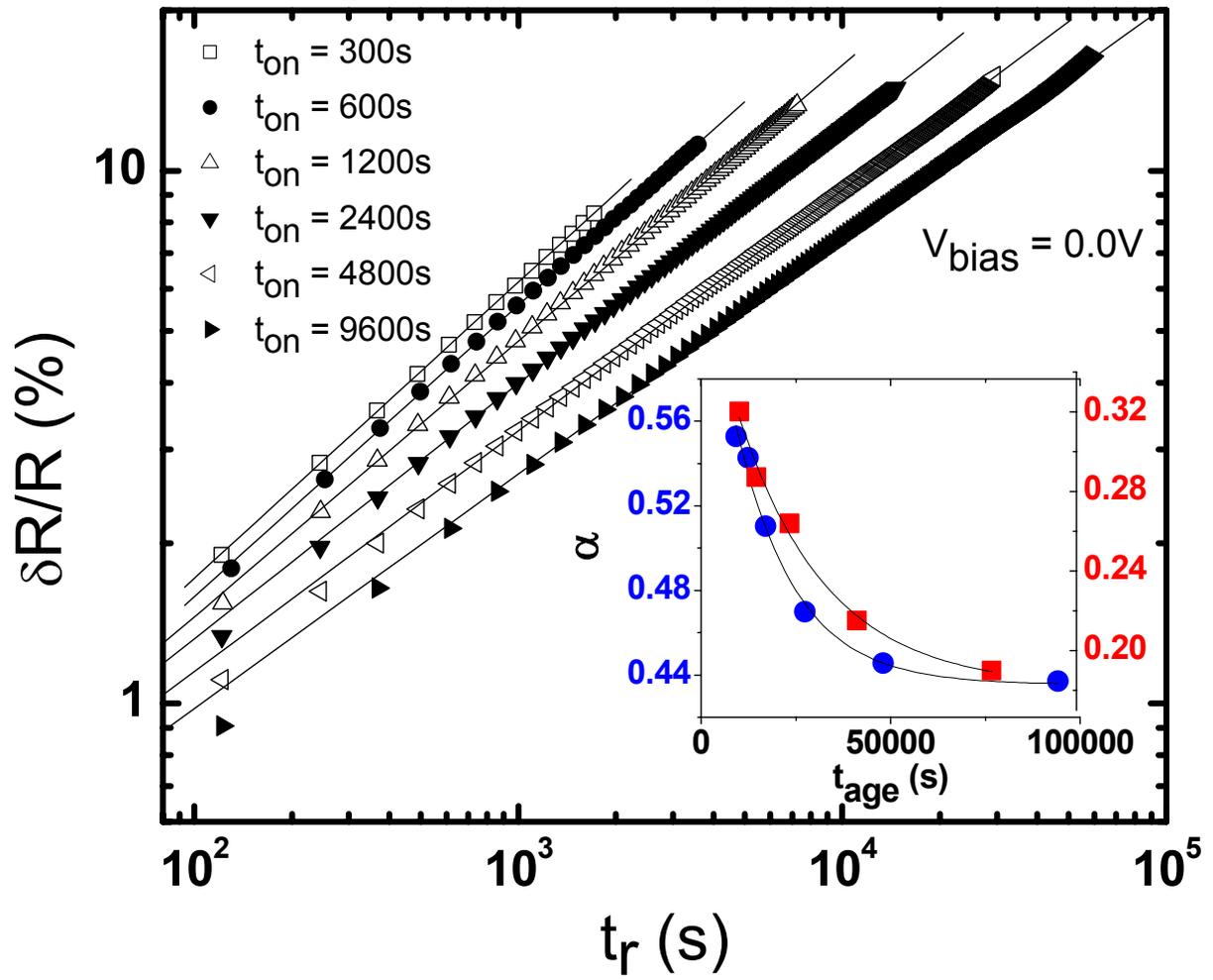

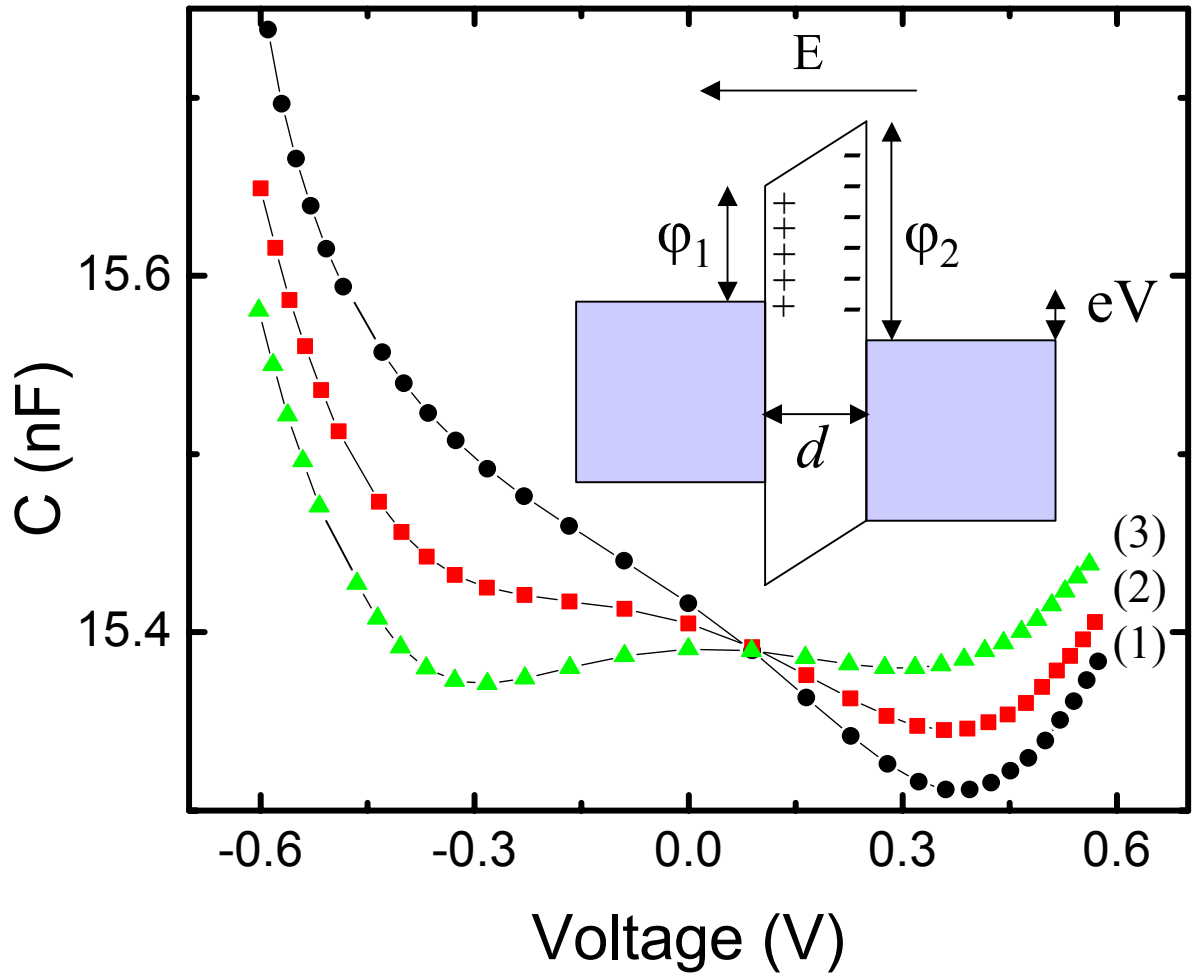

Nesbitt/Figure 4